\documentclass{elsart3p}

\usepackage{graphicx}
\usepackage{amssymb}

\begin{document}

\journal{J. Phys. Chem. Solids}
\date{19 February 2008}

\begin{frontmatter}
\title{Effect of Cd$^{2+}$ on the Growth and Thermal Properties of K$_2$SO$_4$ crystal}
\author[Hadhramout]{S. Bin Anooz\corauthref{cor}},
\corauth[cor]{Corresponding author.}
\ead{s$_{-}$binanooz@yahoo.com}
\author[IKZ]{D. Klimm},
\ead{klimm@ikz-berlin.de}
\author[IKZ]{M. Schmidbauer},
\author[IKZ]{R. Bertram}, and
\author[IKZ]{M. Ro\ss berg}
\address[Hadhramout]{Physics Department, Faculty of Science, Hadhramout University of Science and Technology, Mukalla 50511, Republic of Yemen}
\address[IKZ]{Institute for Crystal Growth, Max-Born-Str. 2, 12489 Berlin, Germany}

\begin{abstract}
Single crystals of pure and Cd$^{2+}$ doped potassium sulfate were grown from aqueous solutions by the slow evaporation technique. From nutrient solutions with a CdSO$_4$ concentration of 4\,wt.\% crystals containing 0.014\,wt.\% dopant concentration could be obtained. The X-ray diffraction patterns of powdered crystals confirmed their crystal structures for both cases. Thermal analysis of pure crystals shows that the $\alpha-\beta$ phase transformation peak around $580^{\,\circ}$C is superimposed with spurious effects, while for Cd$^{2+}$ doped crystals this is not the case. The thermal hysteresis of the phase transition is 8\,K for undoped K$_2$SO$_4$ and is reduced to 3.5\,K for K$_2$SO$_4$:Cd$^{2+}$. Compared to undoped crystals, the optical transmittance of Cd$^{2+}$ doped crystals is higher.
\end{abstract}


\end{frontmatter}

\section{Introduction}

Potassium sulfate K$_2$SO$_4$ crystallizes at room temperature in the orthorhombic olivine type structure and has four formula units per D$^{16}_{2h}=Pnma$ unit cell, with lattice constants $a_0=7.476$\,\AA, $b_0=10.071$\,\AA, and $c_0=5.763$\,\AA\ \cite{McGinnety}. The mineral olivine (Mg,Fe)$_2$SiO$_4$ is a major component of the earth's crust. (Mg,Fe)$_2$SiO$_4$ and its isomorphs A$_2$BO$_4$ (where A = K, NH$_4$, Rb, Cs, B = S, Se) exhibit many interesting physical properties and phase equilibria that were studied repeatedly \cite{Dammak,Gauthier,Khemakhem,Klimm05b}.

Upon heating K$_2$SO$_4$ undergoes a first order transformation at $T_\mathrm{t}\approx580^{\,\circ}$C to a hexagonal structure D$^{4}_{6h}=P6_3/mmc$ with $a_0=5.92$\,\AA\ and $c_0=8.182$\,\AA\ (measured at $640^{\,\circ}$C \cite{Arnold}). In this high-$T$ structure the oxygen positions of the SO$^{2-}_4$ tetrahedra are only partially occupied as a result of rotational disorder. K$_2$SO$_4$ crystals that are grown at room temperature from aqueous solution incorporate OH$^{+}_{3}$ ions. The OH$^{+}_{3}$ concentration decays in the temperature region from $300^{\,\circ}$C to $450^{\,\circ}$C and usually crystals are destroyed by this process \cite{Arnold}.

El-Kabbany \cite{El-Kabbany} has reported a thermal hysteresis for the solid phase transformation with $T_\mathrm{t}=571^{\,\circ}$C at heating and $T_\mathrm{t}=566^{\,\circ}$C at cooling. Electrical conductivity measurements on single crystals have been carried out by Choi et al. \cite{Choi} who found $T_\mathrm{t}=586.9^{\,\circ}$C on heating and $T_\mathrm{t}=581.5^{\,\circ}$C on cooling, with a thermal hysteresis of 5.4\,K. Most of the samples crack near $500\pm30^{\,\circ}$C and show an abrupt drop in electrical conductivity. The hysteresis phenomena, cracking, and electrical ``pretransition phenomena'' were attributed to rotational disorder of SO$_4^{2-}$ ions \cite{Choi} and inclusions of OH$_3^+$ \cite{Arnold}. The present authors reported recently, that the addition of Cd$^{2+}$ to the nutrient solution during K$_2$SO$_4$ crystal growth leads to an improvement of the crystalline quality \cite{BinAnooz}. In this paper, the influence of Cd$^{2+}$ on the growth and on thermal and optical properties of K$_2$SO$_4$ crystal is studied in more detail.

\section{Experimental}

\begin{figure}[htb]
\begin{center}
\includegraphics[height=4.8cm]{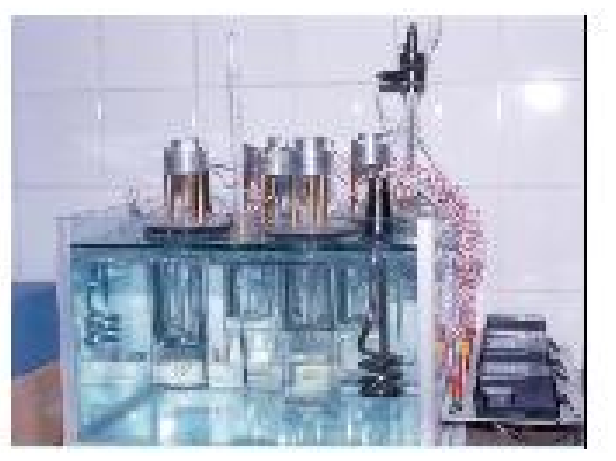} 
\includegraphics[height=4.8cm]{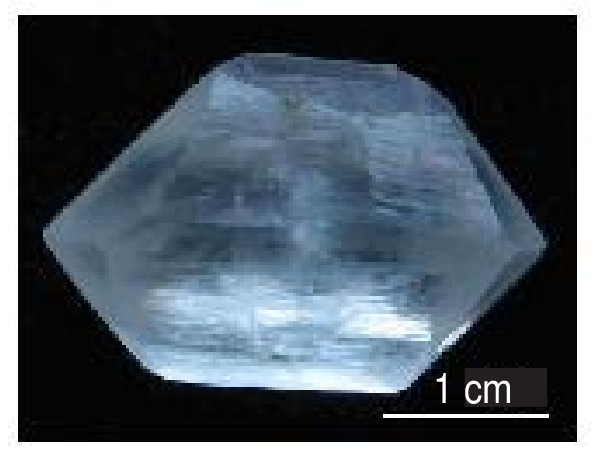}
\end{center}
\caption{top: Multi-jars crystallizer used for the simultaneous growth of undoped and doped K$_2$SO$_4$ crystals. bottom: As grown undoped K$_2$SO$_4$ crystal.}
\label{fig1:exmp}
\end{figure}

Colorless and transparent crystals of pure potassium sulfate could be obtained by slow evaporation of the solvent from saturated aqueous solutions. K$_2$SO$_4$ usually crystallized in prismatic crystals, the $\vec{b}$-axis was found along the long axis of the prism and the $\vec{c}$-axis was along one edge of the quasi-triangular basal plane. K$_2$SO$_4$:Cd$^{2+}$ and K$_2$SO$_4$:Fe$^{2+}$ crystals were grown by the same method from solutions containing 4\,wt.\% of CdSO$_4$ or FeSO$_4$, respectively. The growth process was performed in a multi-jar crystallizer (Fig~\ref{fig1:exmp}, top) to ensure identical growth conditions. In a period of 60~days, we were able to grow colorless, transparent K$_2$SO$_4$ single crystals with well-developed faces having dimensions of $\sim2.5\times1.5\times1$\,cm$^3$ as shown in Fig.~\ref{fig1:exmp}, bottom. Contrary, the growth rate of colorless K$_2$SO$_4$:Cd$^{2+}$ crystals was much lower: in a period of $\sim70$~days crystals up to 3\,mm in diameter could be grown.

The crystals were characterized by Inductively Coupled Plasma--Optical Emission Spectroscopy (ICP--OES), by X-ray powder diffractometry, and by simultaneous Differential Thermal Analysis/Thermogravimetry (DTA/TG), as described recently \cite{BinAnooz}. A NETZSCH STA409CD DTA/TG/QMS (QMS = Quadrupole Mass Spectrometer) system with skimmer coupling of the QMS was used for the simultaneous analysis of gaseous species that are evaporating from the sample during heating \cite{Emmerich}. The shape and position of melting peaks was additionally investigated with a DSC/TG sample carrier and Pt/Rh crucibles with lid. The optical transmission was measured in the region $120-230$\,nm (VUV spectrometer made by Laser Centre Hannover/LZH).

\section{Results and Discussion}

For the undoped K$_2$SO$_4$ crystals only impurities on the ppm level could be found by ICP--OES. If crystal growth was performed from solutions containing 4\,wt.\% CdSO$_4$, the resulting crystals contained only 0.014\,wt.\% of the dopant. From a K$_2$SO$_4$ nutrient solution with 4\,wt.\% FeSO$_4$, crystals with 0.076\,wt.\% Fe concentration were harvested. The small value of the Cd$^{2+}$ distribution coefficient $k_\mathrm{Cd}\approx0.014/4.0=3.5\times10^{-3}$ is in agreement with literature data for the system K$_2$SO$_4$--CdSO$_4$ \cite{Nassau77} that report an almost vanishing solubility. For iron one calculates a larger distribution coefficient $k_\mathrm{Fe}\approx0.076/4.0=19\times10^{-3}$. But this value is still small, and agrees with Christov's observation \cite{Christov04} that almost pure K$_2$SO$_4$ or Na$_2$SO$_4$ are crystallizing from aqueous solutions containing Fe$_2$(SO$_4$)$_3$ or FeSO$_4$, respectively. The small values of $k_\mathrm{Fe}$ and especially $k_\mathrm{Cd}$ must lead to high dopant concentrations in the liquid interface layer and slows down the crystallization process, as compared to the growth of undoped crystals.

From the X-ray diffraction powder patterns of $d$-values and unit cell parameters were calculated with XPowder computer software \cite{Martin}. The cell parameters that are reported in Table~\ref{tab:Xray} match very well with the standard values for K$_2$SO$_4$ as found in the literature (PDF card No. 01-070-1488) \cite{McGinnety}.

\begin{table}[htbp]
\caption {Lattice parameters for pure and doped K$_2$SO$_4$ crystals.}
\begin{center}
\begin{tabular}{lcccc}
\hline
\multicolumn{1}{c}{} & \multicolumn{3}{c}{Lattice constants}\\
\cline{2-4}
Crystal               & $a_0$  [\AA] & $b_0$  [\AA] & $c_0$  [\AA] & Volume [\AA$^3$] \\
\hline
PDF 01-070-1488       & 7.476        & 10.071       & 5.763        & 433.9 \\
K$_2$SO$_4$           & 7.489        & 10.073       & 5.763        & 434.7 \\
K$_2$SO$_4$:Cd$^{2+}$ & 7.505        & 10.064       & 5.766        & 435.5 \\
K$_2$SO$_4$:Fe$^{2+}$ & 7.497        & 10.062       & 5.767        & 435.1 \\
\hline
\end{tabular}
\label{tab:Xray}
\end{center}
\end{table}

\begin{figure}[htb]
\begin{center}
\includegraphics[width=0.4\textwidth]{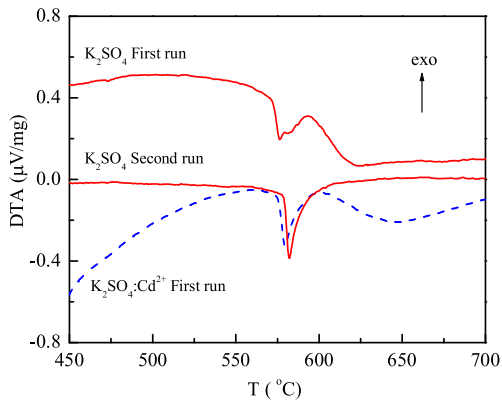}
\end{center}
\caption{DTA heating run with 10\,K/min for K$_2$SO$_4$ and K$_2$SO$_4$:Cd$^{2+}$ crystals.}
\label{fig3:exmp}
\end{figure}

\begin{figure}[htb]
\begin{center}
\includegraphics*[width=0.4\textwidth]{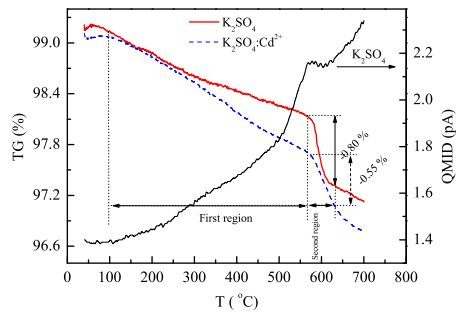}
\end{center}
\caption{Thermogravimetric first heating run for K$_2$SO$_4$ and K$_2$SO$_4$:Cd$^{2+}$ crystals with 10\,K/min and QMID for $m/z=18$ (water).}
\label{fig4:exmp}
\end{figure}

Fig.~\ref{fig3:exmp} compares results of DTA measurements with heating rates of +10\,K/min around $T_\mathrm{t}$ of pure K$_2$SO$_4$ and of K$_2$SO$_4$:Cd$^{2+}$ crystals. The curves reveal the expected endothermic peaks around $580^{\,\circ}$C related to the phase transition of pure and Cd$^{2+}$ doped K$_2$SO$_4$. In the first heating run of pure K$_2$SO$_4$ the phase transition peak is superimposed by other effects while in the second heating run the phase transition peak occurs without superimposed features. It will be shown later that traces of water are incorporated in the (formally anhydrous) crystal structure. This water evaporates during the first heating run, and the second heating curve of pure K$_2$SO$_4$ shows the phase transition peak without additional effects. It is surprising, that K$_2$SO$_4$ doped with Cd$^{2+}$ already in the first heating run did not show the superimposed features. An explanation of this phenomenon can be found using the TG (thermogravimetry) and QMID (Quasi Multiple Ion Detection) curves of K$_2$SO$_4$ and K$_2$SO$_4$:Cd$^{2+}$ crystals that are shown in Fig.~\ref{fig4:exmp}.

During the TG measurement the gas flowing around the K$_2$SO$_4$ sample was analyzed by a QMS system with 298 scans for relative mass numbers $1\leq m/z\leq50$ ($m$ -- molecule mass, $z$ -- charge), this means that one mass spectrum is available every 2.5\,K. From these mass spectra QMID curves can be calculated that show the ion current for every individual $m/z$ within the mass detection range given above. Fig.~\ref{fig4:exmp} shows such QMID curve for the signal $m/z=18$ (H$_2$O) of undoped K$_2$SO$_4$ together with the TG curves. The peak around $T_\mathrm{t}$ proves that evaporating water is the origin of the mass loss. The monotonous slope of the QMID background is typical for skimmer QMS coupling systems and plays no role in this discussion.

The TG curves can be divided into two temperature regions: the first one ranges from $100^{\,\circ}$C up to the onset temperature of the DTA peak ($T\approx580^{\,\circ}$C) and the second region is around the phase transition ($580^{\,\circ}\mathrm{C}\lesssim T\lesssim640^{\,\circ}\mathrm{C}$). In region~I the pure crystal looses about $-1$\% of its initial mass, while for Cd$^{2+}$ doped crystals the mass loss is $-1.36$\%. The smaller mass loss of K$_2$SO$_4$ becomes remarkable for $T>300^{\,\circ}$C were the TG curve for this sample is bent upward.

In region~II the mass loss of K$_2$SO$_4$:Cd$^{2+}$ is smaller than the mass loss of pure K$_2$SO$_4$ by 0.25\%. It is obvious, that the small Cd$^{2+}$ content encourages the evaporation of water already before the onset of the phase transformation. An explanation if this behaviour cannot be given on the basis of the current data. It should be added here that the influence of the $>5$ times larger iron content in K$_2$SO$_4$:Fe$^{2+}$ is just the opposite: Here the TG step at $T_\mathrm{t}$ is larger compared with undoped K$_2$SO$_4$ (not shown in Fig.~\ref{fig4:exmp}).

Arnold et al. \cite{Arnold} reported that crystals of K$_2$SO$_4$ grown at room temperature from aqueous solution incorporate OH$_3^{+}$ ions. They decay in the temperature region from $300^{\,\circ}$C to $450^{\,\circ}$C. If K$_2$SO$_4$ is crystallized above $40^{\,\circ}$C, the content of water in crystals can be minimized \cite{Choi,Chen}. Therefore investigation of the effect of crystal growth temperature in comparison with different concentrations of Cd$^{2+}$ on this phenomena of K$_2$SO$_4$ crystals will be considered in future studies.

\begin{figure}[htb]
\begin{center}
\includegraphics*[width=0.4\textwidth]{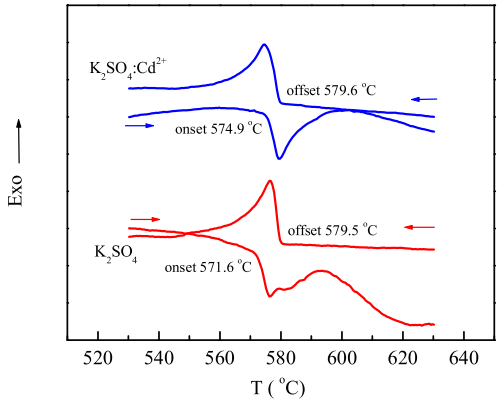}
\end{center}
\caption{DTA first heating and cooling curves for K$_2$SO$_4$:Cd$^{2+}$ and K$_2$SO$_4$ crystals.}
\label{fig5:exmp}
\end{figure}

DTA heating and cooling curves around the phase transition peak for K$_2$SO$_4$ and K$_2$SO$_4$:Cd$^{2+}$ crystals are shown in Fig.~\ref{fig5:exmp}. The transition temperature $T_\mathrm{t}$ can be determined from the intersection of the extended basis line with the tangent at the inflection point (extrapolated onset). It is remarkable that for pure K$_2$SO$_4$, $T_\mathrm{t}$ as obtained from the first heating or cooling runs, respectively, differ considerably by a hysteresis of about 8\,K. For K$_2$SO$_4$:Cd$^{2+}$ this hysteresis is only 4.5\,K. It was reported recently that the hysteresis is reduced considerably for both K$_2$SO$_4$ and K$_2$SO$_4$:Cd$^{2+}$ in subsequent (starting from the second) heating/cooling runs \cite{BinAnooz}. Hence it can be suggested that the effect of Cd$^{2+}$ on hysteresis is similar to the effect of annealing for K$_2$SO$_4$ crystal, if the first heating/cooling run is considered as annealing for the undoped crystal.

\begin{figure}[htb]
\begin{center}
\includegraphics*[width=0.4\textwidth]{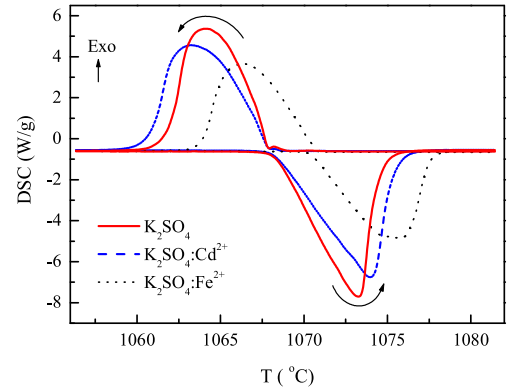}
\end{center}
\caption{DSC heating and cooling curves ($\pm10$\,K/min) near the melting temperature for undoped K$_2$SO$_4$ and for crystals doped with Cd or Fe, respectively.}
\label{fig:melt}
\end{figure}

Fig.~\ref{fig:melt} compares the melting behavior of pure K$_2$SO$_4$ with that of K$_2$SO$_4$:Cd$^{2+}$ and, for reference, with that of a higher doped K$_2$SO$_4$:Fe$^{2+}$. For K$_2$SO$_4$ as well K$_2$SO$_4$:Cd$^{2+}$ the melting temperature $T_\mathrm{f}=(1068.2\pm0.2)^{\,\circ}$C and the crystallization temperature are almost identical with negligible hysteresis. In contrast, the higher dopant concentration in K$_2$SO$_4$:Fe$^{2+}$ leads to remarkable hysteresis of 0.4\,K and shifts the melting point to $T_\mathrm{f}=(1070.4\pm0.1)^{\,\circ}$C. In analogy to Fig.~6 of \cite{BinAnooz} which represents a model for the $\beta/\alpha$ phase transformation, an extended two phase region $\alpha$/\textit{liquid} can explain the shift of $T_\mathrm{f}$ and the hysteresis.

\begin{figure}[htb]
\begin{center}
\includegraphics[width=0.4\textwidth]{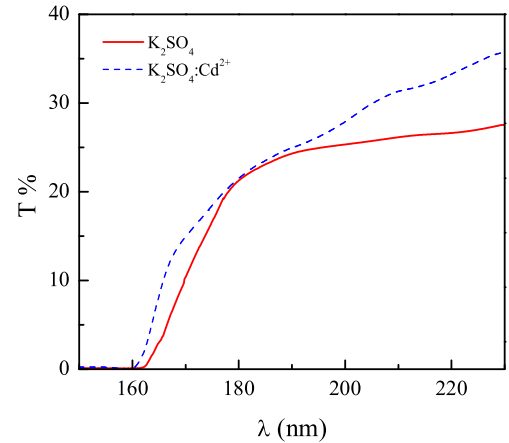}
\end{center}
\caption{Optical transmittance T as a function of wavelength $\lambda$ for K$_2$SO$_4$ and K$_2$SO$_4$:Cd$^{2+}$ crystals.}
\label{fig6:exmp}
\end{figure}

The widths of the K$_2$SO$_4$:Cd$^{2+}$ melting and crystallization peaks are slightly larger compared to the undoped K$_2$SO$_4$, as a result of the minor Cd$^{2+}$ dopant concentration. It is remarkable, however, that this impurity leads in the end to improved optical properties.

Fig.~\ref{fig6:exmp} shows that the optical transmission of K$_2$SO$_4$:Cd$^{2+}$ is higher then the transmission  of K$_2$SO$_4$ for all wavelengths shown here ($\lambda\leq230$\,nm). This may partially be a result of scattering centers in the crystals that were grown from solutions containing Cd$^{2+}$, as Cd$^{2+}$ doped samples can already with the naked eye be recognized to possess a higher optical clarity. Nevertheless it is remarkable that the absorption edge is shifted toward the VUV region, resulting in an increased energy gap of K$_2$SO$_4$:Cd$^{2+}$ compared with undoped K$_2$SO$_4$.

\section{Conclusion}

Pure K$_2$SO$_4$ and Cd doped K$_2$SO$_4$ crystals can be grown from aqueous solution by the slow evaporation technique. The incorporation of Cd is weak: From solutions with 4\,wt.\%\ CdSO$_4$ one obtains Cd$^{2+}$:K$_2$SO$_4$ crystals with only 0.014\,wt.\% of the dopant. Nevertheless, the influence of cadmium is remarkable: Cd doped crystals contain less water, and the remaining water is loosely bond compared with water in undoped crystals. The optical transparency of Cd$^{2+}$:K$_2$SO$_4$ in the UV region is larger.

\ack{

S. Bin Anooz was awarded a scholarship from the ``Germany Academic Exchange Service" (DAAD) for this work that is gratefully acknowledged.}

\end{document}